\documentclass[]{IEEEphot}

\begin{document}

\title{Electro-mechanically induced GHz rate optical frequency modulation in silicon}

\author{S.~Tallur,~\IEEEmembership{Student Member,~IEEE}, S.~A.~Bhave,~\IEEEmembership{Member,~IEEE}}

\affil{OxideMEMS Lab, Department of Electrical and Computer Engineering, \\Cornell
University, Ithaca NY 14850 USA}  


\maketitle

\markboth{IEEE Photonics Journal}{Electro-mechanically induced GHz rate optical frequency modulation in silicon}


\begin{abstract}
We present a monolithic silicon acousto-optic frequency modulator (AOFM) operating at 1.09GHz. Direct spectroscopy of the modulated laser power shows asymmetric sidebands which indicate coincident amplitude modulation and frequency modulation. Employing mechanical levers to enhance displacement of the optical resonator resulted in greater than 67X improvement in the opto-mechanical frequency modulation factor over earlier reported numbers for silicon nanobeams.
\end{abstract}


\section{Introduction}

Optical frequency modulation has been achieved in microstructures using surface acoustic waves to strain self-assembled InAs/GaAs quantum dots \cite{qd1, qd2}. Photon generation at $>$GHz frequency spacing from a single pump laser has been realised in silicon nitride and silicon dioxide via nonlinear optical processes such as the optical Kerr effect \cite{comb} and stimulated Brillouin scattering \cite{sbs1, sbs2}. It is difficult to exploit such non-linear optical phenomena in silicon, primarily owing to two-photon absorption that limits optical power handling. Wavelength conversion in silicon has been demonstrated using fundamentally all-optical schemes \cite{Siopt1, Siopt2} and via free-carrier plasma dispersion effect \cite{lipson}, making their integration on micro-electronic chips challenging.

Electrostatic capacitive actuation and detection is the main scheme of transduction used in Radio Frequency (RF) Micro Electro Mechanical Systems (MEMS) resonators \cite{relec1, relec2, relec3}. The use of electrostatics enables direct integration with electronics used for processing RF signals.  Electrostatic capacitive actuation of opto-mechanical resonators has been demonstrated earlier using a standalone opto-mechanical system with a gradient electrical force on a dielectric provided by an off-chip electrode \cite{offchip}, and using integrated electrodes \cite{onchip1, onchip2, onchip3}. Employing this scheme at GHz rates will provide valuable narrow-band acousto-optic modulators for direct conversion of electrical signals to optical intensity modulation that are valuable towards realising a chip-scale opto-electronic oscillator (OEO) \cite{oeo1}. In addition to intensity modulation, it is desirable to achieve frequency modulation, which adds variable group delay into the optical field, thus potentially boosting phase noise performance of the OEO \cite{oeo2}. In this work, we present a monolithic integrated silicon acousto-optic frequency modulator designed in a silicon-on-insulator (SOI) platform. The modulator bandwidth is governed by the quality factor of the mechanical resonance. This design can open up avenues towards realizing a truly chip-scale OEO and enable on-chip Dense Wavelength Division Multiplexing (DWDM) using a single input laser. The next section presents the theoretical basis for the operation of this device. In later sections, we describe the device design and experimental results.

\begin{figure}[t]
\centering
\includegraphics[width=3in]{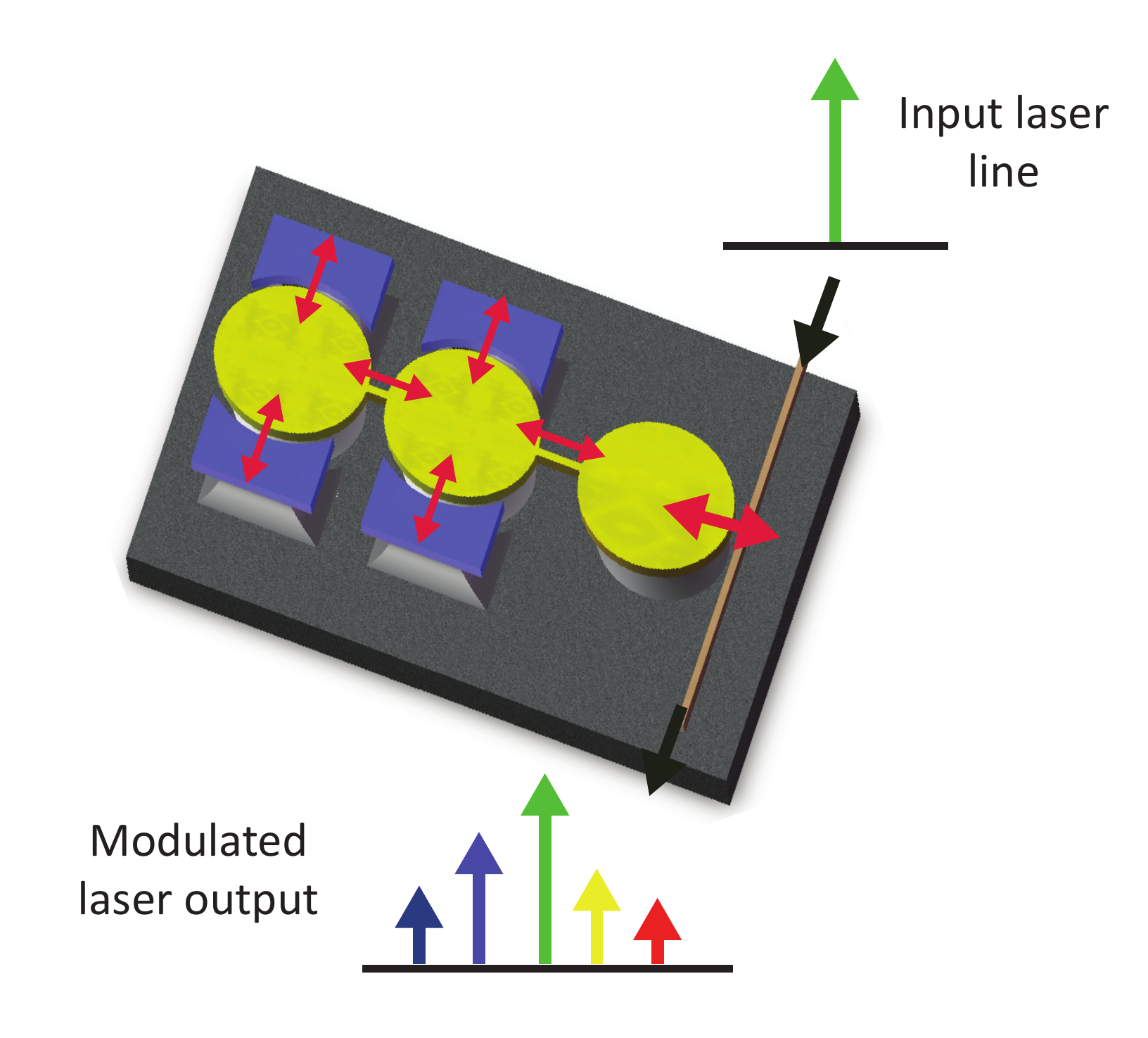}
\caption{Illustration of the acousto-optic frequency modulator showing generation of laser sidebands by acoustically exciting motion of the coupled opto-mechanical resonator via electrostatic capacitive transduction. The motion of the opto-mechanical cavity results in phase modulation of the light field resulting in generation of multiple sidebands around the laser line.}
\label{Fig_illustration}
\end{figure}

\section{Theoretical Treatment Of Opto-Mechanical Modulation}

Silicon opto-mechanical resonators operated in the resolved sideband regime stand out as strong candidates for optical frequency modulation (FM) owing to their strongly coupled mechanical and optical degrees of freedom \cite{kippvahala}. This coupling mechanism is nonlinear and the non-adiabatic response of the intra-cavity optical field to changes in the cavity size can lead to modification of its mechanical dynamics. On the other hand, if the mechanical resonant mode of the cavity is actuated, the cavity executes oscillatory motion at this frequency which results in Doppler-shift of the circulating intra-cavity optical field, thus modulating its phase. In the frequency domain, this frequency modulation manifests itself as sidebands centered about the input laser frequency line, at a frequency separation that is equal to the frequency of the actuated mechanical resonance. The cavity density of states shapes the resultant sidebands leading to enhancement of only those sidebands that are coupled to the optical cavity, thus causing an asymmetry in the intensity of the lower and higher frequency sidebands. This asymmetry leads to an effective amplitude modulation (AM) of the laser light, in addition to frequency modulation. Figure \ref{Fig_illustration} shows an illustration of such a scheme, using electrostatic actuation. This section deals with the theoretical treatment of this phenomenon.

\subsection{Electrostatically induced opto-mechanical modulation}

The microdisk resonators on the left of Figure \ref{Fig_illustration} are flanked by electrodes to provide electrostatic actuation to drive the mechanical modes. Applying a combination of DC bias voltage, $V_{DC}$, and AC signal, $v_{AC}$, results in generation of a force acting on the disk in the radial direction, at the frequency of the applied AC voltage, $\omega_{AC}$:

\begin{equation}
\label{force}
F(\omega_{AC})=\frac{dC_g}{dg}V_{DC}v_{AC}
\end{equation}
where $\frac{dC_g}{dg}$ is the change in the actuation capacitance, $C_g$, for a small change in the electrode-resonator capacitor gap, $g$. If the frequency of the applied AC voltage corresponds to the frequency of a mechanical mode of the resonator, the force results in a radial displacement at the mechanical resonantor frequency as shown in equation \ref{displeqn} below.

\begin{equation}
\label{displeqn}
\Delta r(\omega_{AC})=\frac{Q_{mech}F(\omega_{AC})}{K_r}
\end{equation}

Here $Q_{mech}$ denotes the quality factor of the mechanical resonance and $K_r$ represents the mechanical stiffness (effective spring constant) of the radial vibrational mechanical mode. These vibrations are mechanically coupled into the opto-mechanical cavity resonator on the right, which is probed by coupling laser light to an optical resonant mode of the cavity. The sensitivity of the opto-mechanical resonator to radial displacement is obtained from the optical resonance condition:

\begin{equation}
\label{opt_res}
m\lambda_{opt}=2\pi Rn_{eff}
\end{equation}
where $m$ is an integer, $\lambda_{opt}$ is the free space wavelength of the optical resonant mode, $R$ is the radius of the opto-mechanical resonator and $n_{eff}$ is the effective refractive index for the mode. A displacement $\Delta r(\omega_{AC})$ results in a shift in optical resonance wavelength of $\Delta \lambda_{opt}=\frac{\Delta r(\omega_{AC})}{R}\lambda_{opt}$. This can also be quantified by defining an equivalent opto-mechanical coupling coefficient:

\begin{equation}
\label{opt_res}
g_{om}=\frac{d\omega_{opt}}{dR}=-\frac{d\lambda_{opt}}{dR}
\end{equation}
where $\omega_{opt}$ is the frequency of the input pump laser. The mechanical vibration of the resonator is detected as modulation of the input laser light, as detailed in the following sub-section.

\subsection{Transmission of an opto-mechanical cavity}

Assume that the cavity is oscillating at the mechanical resonant frequency, $\Omega_0$, such that its radius is sinusoidally displaced as described in equation \ref{cavity_displ}:

\begin{equation}
\label{cavity_displ}
u(t)=U_0sin(\Omega_0t)
\end{equation}

The amplitude of the light field circulating inside the opto-mechanical cavity, $a_p(t)$, can be derived as in \cite{ressb} by introducing a modulation index, $\beta=\frac{U_0}{R}\frac{\omega_{opt}}{\Omega_0}$:

\begin{equation}
\label{intracav_field}
a_p(t)=\frac{s}{\sqrt{\tau_{ex}}}\sum_{n=-\infty}^{+\infty}\frac{(-i)^nJ_n(\beta)}{\frac{\kappa}{2}+i(\Delta+n\Omega_0)}exp[i(\omega_{opt}+n\Omega_0)t+i\beta cos(\Omega_0t)]
\end{equation}
where $s$ denotes the input pump laser field amplitude, $\tau_{ex}$ is the photon lifetime due to coupling between the cavity and the waveguide, $\kappa$ is the full width at half maximum (FWHM) linewidth of the optical cavity resonance, and $\Delta$ denotes the detuning of the laser from the optical cavity resonance, specified in angular frequency. Thus, the modulation index determines the relative weights of the sidebands, whose width is governed by the optical cavity linewidth.

Of particular interest are the cases where $\kappa/2\gg(\Delta+\Omega_0)$ and $\kappa/2\ll(\Delta+\Omega_0)$. The former case corresponds to unresolved motional sidebands. Noting that  $J_{-n}(\beta)=(-1)^nJ_n(\beta)$, it is easy to see that the lower and higher frequency primary sidebands are in phase with each other, resulting in predominant amplitude modulation. In the latter case of resolved motional sidebands, the phase relationship is reversed, leading to predominant frequency modulation.  For other detuning values, the sidebands are highly asymmetric in the resolved sideband regime. When the detuning is zero, the motion of the opto-mechanical cavity leads to pure phase modulation of the laser light. For non-zero detuning, the cavity density of states leads to an asymmetry in the relative sideband intensities. This asymmetry is associated with coincidental amplitude modulation and frequency modulation of the laser light \cite{roder}, as explained in the following sub-section.

\subsection{Coincidental amplitude and frequency modulation}

Consider a sine wave carrier expressed by the general equation below:

\begin{equation}
\label{sine_wave}
s(t)=Acos(\omega_ct)
\end{equation}

$A$ is the amplitude and $\omega_c$ is the frequency of the carrier signal. When the signal is amplitude-modulated with a modulation index $m_{AM}$ at frequency $\omega_m$, the expression for the modulated signal can be written as follows:

\setlength{\arraycolsep}{0.0em}
\begin{eqnarray}
\label{AM_wave}
s_{AM}(t)&{}={}&A[1+m_{AM}cos(\omega_mt)]cos(\omega_ct)\nonumber\\
&&=Acos(\omega_ct)+\frac{m_{AM}A}{2}cos[(\omega_c+\omega_m)t]+\frac{m_{AM}A}{2}cos[(\omega_c-\omega_m)t]
\end{eqnarray}
\setlength{\arraycolsep}{5pt}

Thus, amplitude modulation generates symmetric sidebands around the carrier at frequencies $\omega_c+\omega_m$ and $\omega_c-\omega_m$. For 100$\%$ modulation, i.e. $m_{AM}$ = 1, the amplitude of each band is half that of the carrier, which corresponds to 6dB lower power than carrier. In amplitude modulation, all the power injected into the signal via the modulating signal is added at the sidebands, without affecting power at the carrier frequency.

Consider frequency-modulation of the carrier at frequency $\omega_m$ with a modulation index $m_{FM}$. The modulated signal now has multiple frequency components:

\setlength{\arraycolsep}{0.0em}
\begin{eqnarray}
\label{FM_wave}
s_{FM}(t)&{}={}&A[cos(\omega_ct+m_{FM}sin(\omega_mt))]\nonumber\\
&&=A[cos(\omega_ct)cos(m_{FM}sin(\omega_mt))-sin(\omega_ct)sin(m_{FM}sin(\omega_mt))]
\end{eqnarray}
\setlength{\arraycolsep}{5pt}

Using the Fourier series expansions $cos(m_{FM}sin(\omega_mt))=\sum{}_{n=0}^\infty 2nJ_{2n}(m_{FM})cos(2n\omega_mt)$ and $sin(m_{FM}sin(\omega_mt))=\sum{}_{n=0}^\infty 2J_{2n+1}(m_{FM})sin([2n+1]\omega_mt)$, we can express the signal in terms of all its frequency components:

\setlength{\arraycolsep}{0.0em}
\begin{eqnarray}
\label{FM_sidebands}
s_{FM}(t)&{}={}&A[J_0(m_{FM})cos(\omega_ct)\nonumber\\
&&+\sum{}_{n=0}^\infty (-1)^nJ_n(m_{FM})[cos((\omega_c-n\omega_m)t)+(-1)^ncos((\omega_c+n\omega_m)t)]]
\end{eqnarray}
\setlength{\arraycolsep}{5pt}

The expression highlights the phase relation between all the frequency components. Each sideband is characterised by its order $n$. The vector sums of the odd-order sideband pairs are always in quadrature with the carrier component of the signal and the vector sums of the even-order sideband pairs are always collinear with the carrier component. The phase shift for the odd-order sidebands arises from the reversal of phase for the lower frequency sidebands with respect to the higher frequency sidebands. The power in the lower and higher frequency sidebands is the same for sidebands of all order. Unlike AM, FM affects the power at the carrier frequency. FM does not involve injecting power into the signal - the modulating signal results in redistribution of the power at the carrier frequency into all the sidebands.
From equations \ref{AM_wave} and \ref{FM_sidebands}, it is clear that if a signal has coincidental AM and FM, the higher and lower frequency sidebands of the first order are asymmetric, i.e. the power in the higher frequency sideband is different from the power in the lower frequency sideband. The mismatch in power depends on the phase relationship between the AM and FM processes. It is also interesting to note the effect of introducing asymmetry in the intensities of lower and higher frequency sidebands in equation \ref{FM_sidebands}. The amplitude of the resulting signal is no longer constant, and has AM in addition to FM.

\section{Device Design}

\subsection{Mechanical displacement amplification}

A silicon acousto-optic modulator (AOM) with a mechanical resonance frequency of 1.1GHz has been presented \cite{sureshmems}. This device has an optical cavity 3dB-linewidth $\approx$5.88GHz and results in predominant amplitude modulation of the laser light. For efficient frequency modulation we want the device to operate in the resolved sideband regime. If the laser frequency is detuned to the full-width-half-maximum (FWHM) point of the optical resonance, $\omega_{opt}$, the modulation depth depends linearly on the motion induced frequency shift. The opto-mechanical coupling coefficient for a micro-ring resonator is given by  $g_{om}=\frac{d\omega_{opt}}{dR} =-\frac{\omega_{opt}}{R}$, where $R$ is the radius of the micro-ring. Thus, larger mechanical displacement translates to greater shift in the resonance frequency, thereby enhancing the modulation depth. Hence it is desirable to get large mechanical displacements in addition to operating in the resolved sideband regime, to achieve efficient frequency modulation.

\begin{figure}[t]
\centering
\includegraphics[width=5.5in]{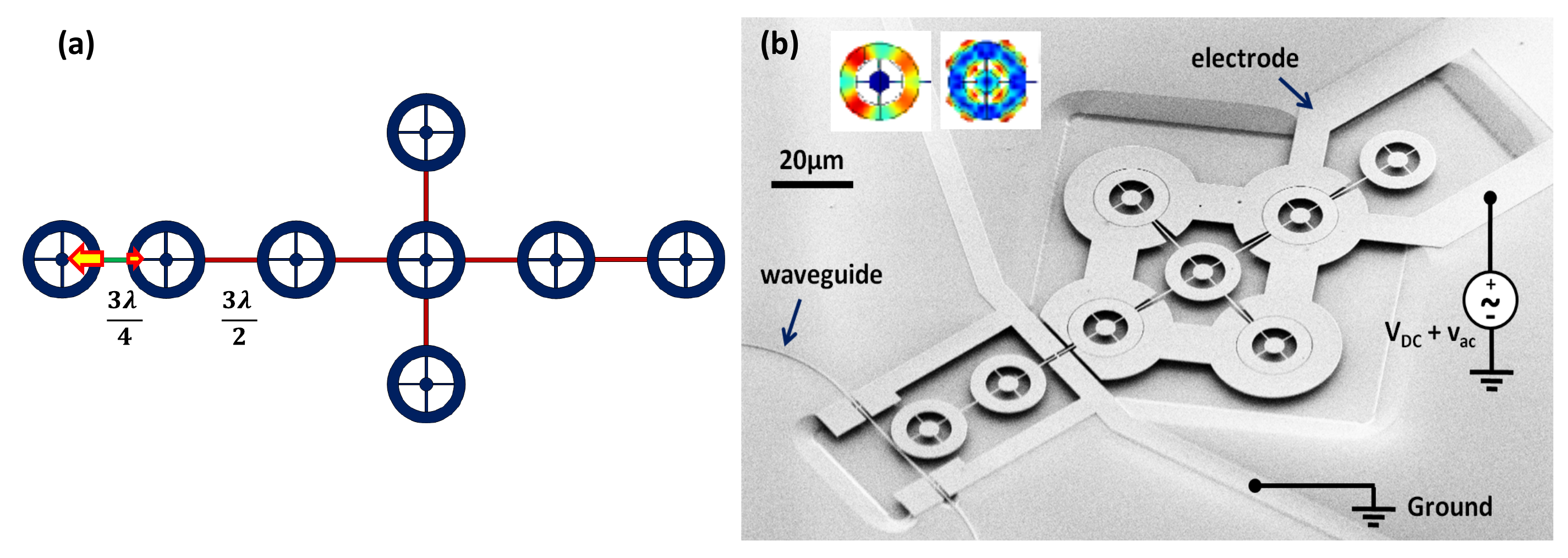}
\caption{(a) Illustration of the micromechanical displacement amplifier. The three-quarter wavelength coupling beam connecting the output opto-mechanical resonator on the left to the array composite of drive micro-mechanical resonators on the right ensures mechanical displacement amplification. (b) Scanning electron micrograph (SEM) of the fabricated acousto-optic frequency modulator (AOFM). Insets: Finite element simulation (FEM) of the mechanical mode shape of the ring at 176MHz (left - fundamental radial expansion mode) and 1.09GHz (right - compound radial expansion mode).}
\label{array}
\end{figure}

The modulator presented in \cite{sureshmems} is extended to an array of mechanically coupled ring resonators for large electrostatic driving force. We also use a mechanical lever system \cite{displamp} to realize displacement amplification. An array composite of rings on the input electrostatic side feeds via an acoustic quarter-wavelength coupling beam into a single output opto-mechanical ring resonator, as illustrated in Figure \ref{array}.a. 

Consider a strongly coupled array of $N$ ring resonators strongly coupled to each other using half-wavelength coupling beams. The effective stiffness of the composite resonator defined by the array is $K_{eff}=NK_r$, where $K_r$ is the stiffness of each individual ring. Coupling this array to another ring resonator using a quarter-wavelength coupling beam results in a mismatch of the stiffnesses on either side of this coupling beam. A quarter-wavelength coupling beam, by definition, forces the velocity at both ends of the beam to be the the same. Thus, the kinetic energy of the moving ends of the coupling beam, and hence the stored potential energy in the resonators on either end of the beam is the same. Thus, if $x_o$ is the radial displacement of the output opto-mechanical ring resonator, and $x_{array}$ is the radial displacement of each ring resonator in the array, the energy constraint ensures:

\setlength{\arraycolsep}{0.0em}
\begin{eqnarray}
\label{displamp}
\frac{1}{2}K_rx_o^2=\frac{1}{2}K_{eff}x_{array}^2\\
\Rightarrow\frac{x_o}{x_{array}}=\sqrt{N}
\end{eqnarray}
\setlength{\arraycolsep}{5pt}

Thus, the difference in stiffnesses of the input array and output ring results in an enhancement of the displacement at the output\footnote{The coupling beam lengths are designed to correspond to three-halves of the acoustic wavelength for strong coupling, and three-quarters of the acoustic wavelength at 1.09GHz for the displacement amplification beam. These dimensions are chosen for making the layout of the rings feasible. Choosing such a geometry does not affect the displacement amplification scheme.}.

\begin{figure}[t]
\centering
\includegraphics[width=5.5in]{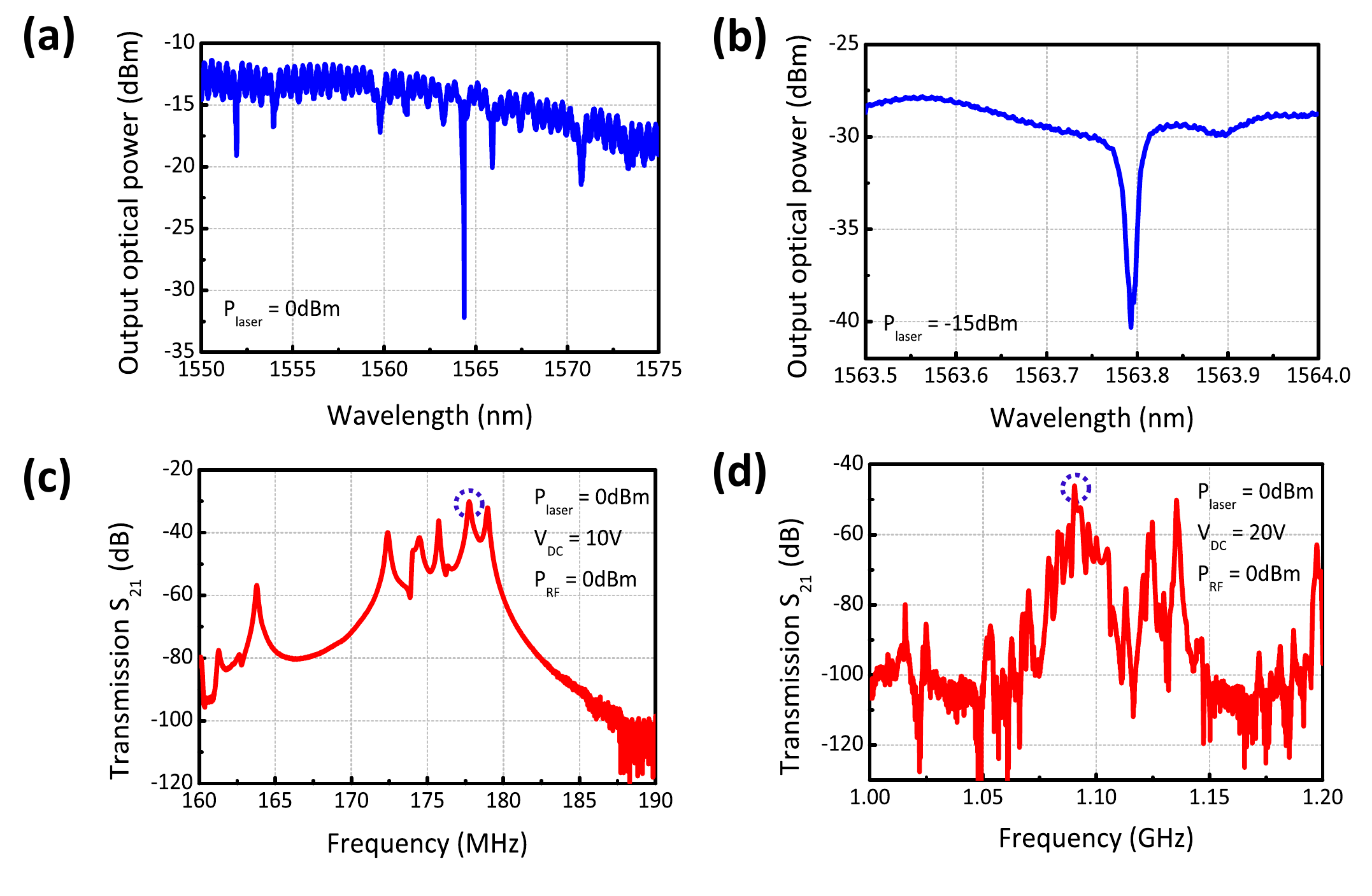}
\caption{(a) Various resonances for the opto-mechanical cavity are observed in the optical transmission spectrum as the laser wavelength is swept from 1,550nm to 1,575nm. (b) Critically coupled optical resonance at 1,563.8nm with loaded optical Q $\approx$ 150,000, probed with an input laser power of -15dBm. (c) Transmission of the modulator measured from 160MHz to 190MHz. The fundamental radial expansion mode of the ring has an insertion loss of 30dB, seen here in the transmission spectrum (circled). (d) Transmission of the modulator measured from 1GHz to 1.2GHz. The compound radial expansion mode of the ring has an insertion loss of 46dB (circled). Multiple peaks are seen as a result of imperfections in the beam lengths on account of fabrication variations. All experiments were performed under atmospheric pressure at room temperature.}
\label{charac}
\end{figure}

\subsection{Fabrication process}

This section describes how the modulator is fabricated. The procedure involves a three mask process flow on a custom silicon-on-insulator (SOI) wafer (undoped 250nm device layer for low optical loss and 3$\mu$m thick buried oxide for isolation of the device from the silicon substrate). The top silicon is thermally oxidised to obtain a thin oxide hard mask layer of thickness 60nm atop a 220nm thick silicon device layer. ma-N 2403 electron beam resist is spun on top of the oxide and patterned using electron beam lithography. The patterns are transferred into the silicon dioxide using a CHF$_3$/O$_2$ based reactive ion etcher and then into the silicon device layer using a chlorine based reactive ion etch. A layer of SPR-220 3.0 photoresist is spun and a second mask is used to pattern windows above the mechanical resonator, the electrical routing beams and the bond-pads. This is followed by a boron ion implantation and nitrogen anneal to reduce the resistivity of these structures. A third mask is then used to pattern release windows near the modulator using SPR-220 3.0 photoresist, followed by a timed release etch in buffered oxide etchant to undercut the devices. The samples are then dried using a critical point dryer to prevent stiction. Figure \ref{array}.b shows a scanning electron micrograph (SEM) of the fabricated device.

\section{Experimental Setup and Results}

\subsection{Characterization of the modulator}

To probe the optical response of the modulator, light from a Santec TSL-510 tunable laser polarized parallel to the substrate is coupled into the waveguide using grating couplers. The optical output power is measured to obtain the optical transmission spectrum of the device. As shown in Figure \ref{charac}.a we observe multiple dips corresponding to various modes of the optical cavity. We choose the critically coupled optical resonant mode at a wavelength around 1,564nm with a total optical quality factor of $\approx$ 150,000 and extinction 10dB, as seen in Figure \ref{charac}.b. This corresponds to a full-width half maximum (FWHM) cavity linewidth of 1.27GHz.
To identify the mechanical modes of the resonator, we characterize the RF transmission of the response of the modulator using a network analyzer. A bias-tee is used to apply a combination of DC bias voltage along with AC power from port 1 of an Agilent N5230A network analyzer at the probe-pads of the electrode with a Cascade GSG RF probe. The tunable laser is blue detuned to the half maximum point of the optical cavity. Mechanical motion of the device induced by the applied input voltage translates into intensity modulation of the laser light, which is picked off and converted into electrical RF signal by a Newport 1544-A near-IR photoreceiver connected at the output of the modulator. The output of the photoreceiver is connected to port 2 of the network analyzer. The RF frequency is swept and the transmission spectrum for the modulator is recorded as shown in Figures \ref{charac}.c and \ref{charac}.d. The mechanical quality factors measured for these resonant modes are 800 and 1,800 respectively. The coupling beams are designed for optimal displacement amplification of the mechanical mode with frequency 1.09GHz.

\subsection{Characterization of the optical-sidebands}

\begin{figure}[t]
\centering
\includegraphics[width=4in]{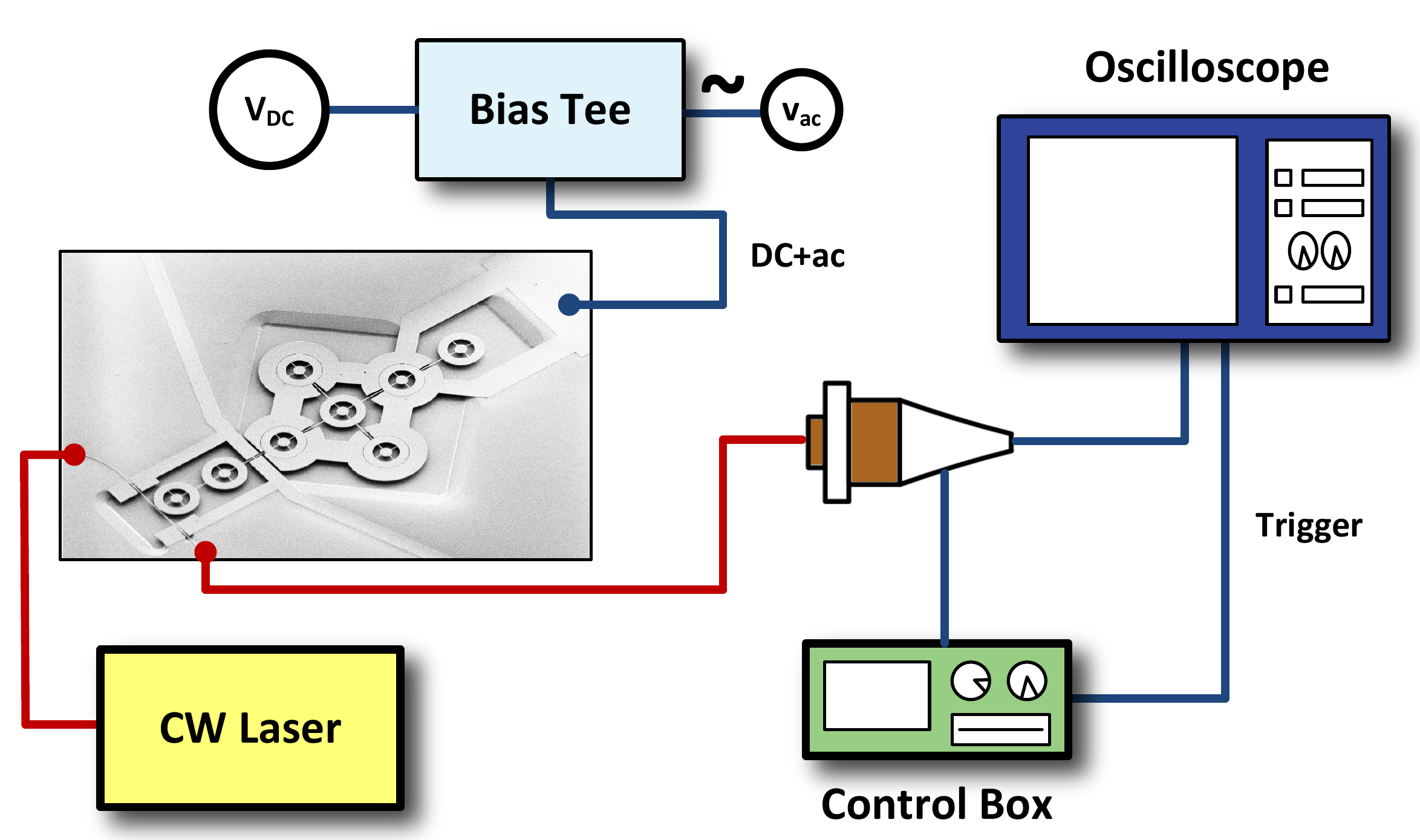}
\caption{Experimental setup for characterization of the optical sidebands of the modulator using a scanning FP interferometer. The frequency of the applied AC voltage corresponds to the resonance frequency of the mechanical mode of interest (circled in Figures \ref{charac}.c and \ref{charac}.d).}
\label{FP_setup}
\end{figure}

For observing the motion-generated optical sidebands for the modulator, we use a Thorlabs SA210-12B scanning Fabry-Perot (FP) interferometer. The confocal design of this FP Interferometer cavity is relatively insensitive to the alignment of the input beam. The tunable laser is blue detuned to the optical cavity and a bias-tee is used to apply a combination of DC bias voltage and AC voltage at the resonance frequency of the mechanical mode (circled in Figures \ref{charac}.c and \ref{charac}.d) using an Agilent E8257D PSG Analog Signal Generator. Using an SA201 control box for the FP interferometer, the wavelength of the FP cavity is scanned across its entire range by sweeping the voltage on the piezo-controller in the control box. The transmitted light intensity is measured using an internal photodiode, amplified by a transimpedance amplifier inside the control box, and displayed on an oscilloscope. As the FP cavity scans across wavelengths, its output on an oscilloscope shows a peak whenever it passes across a sideband. The controller also provides a trigger signal to the oscilloscope, which allows the oscilloscope to easily trigger at the beginning or the middle of the scan. The free spectral range (FSR) of the FP interferometer that we use is 1.5GHz, and this information is used to scale the x-axis of the oscilloscope output from units of time to frequency offset from pump laser. The root mean square (RMS) noise voltage of the photo-amplifier in the control box is 1.5mV for a gain setting of 1MV/A. The measured FSR for the cavity is 7.2ms and the full width half maximum (FWHM) linewidth is 35$\mu$s, which correspond to 1.5GHz and 7.2MHz respectively, and cavity finesse of 205. An illustration of the experimental setup is shown in Figure \ref{FP_setup}.

\begin{figure}[t]
\centering
\includegraphics[width=4in]{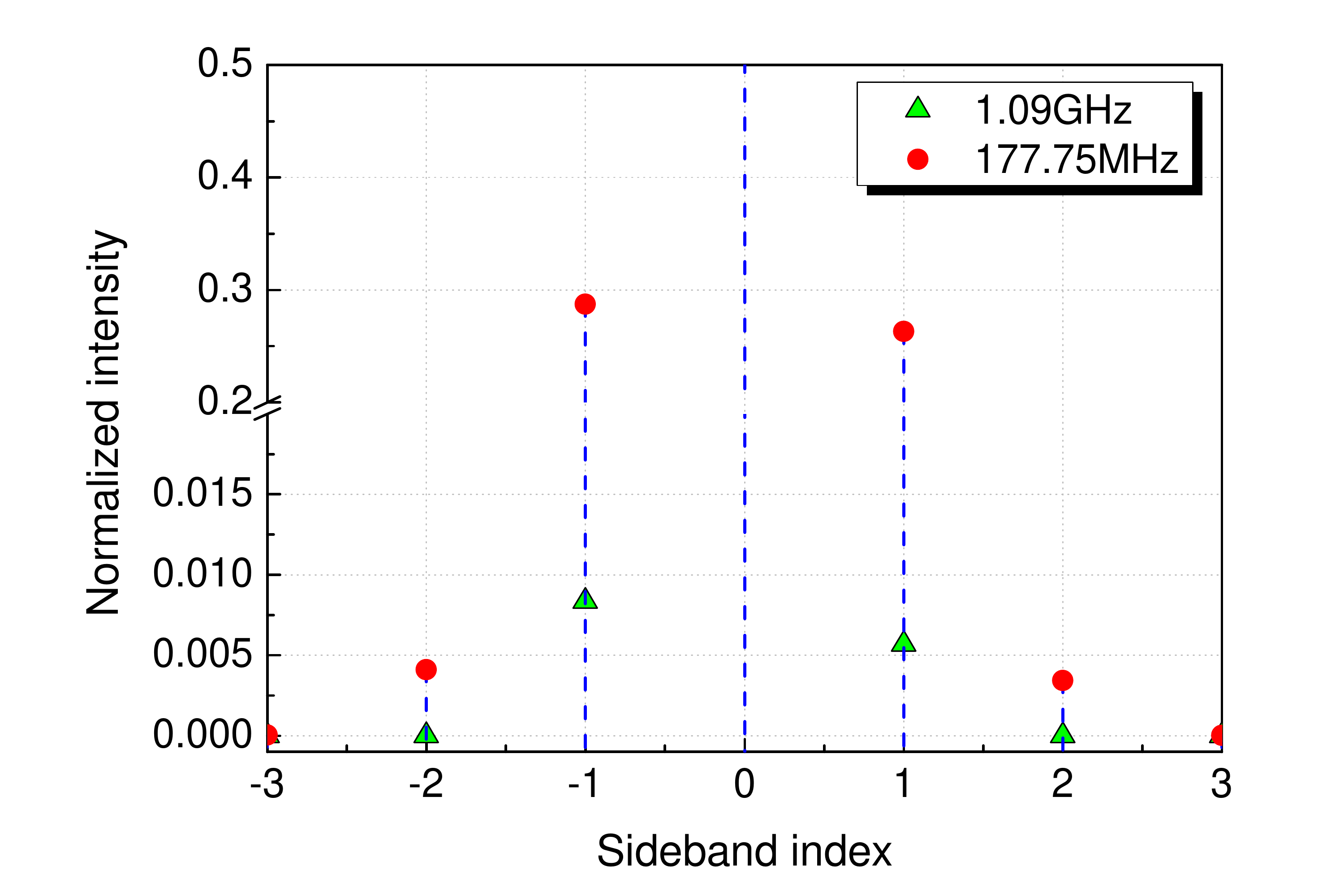}
\caption{Calculated spectrum for the modulator response at 177.75MHz and 1.09GHz for relative detuning of 0.5. A combination of 10V DC and 5dBm AC power was considered at the input electrode for operation at 177.75MHz, and a combination of 20V DC and 8dBm AC power was accounted for at 1.09GHz. The primary sideband intensities are highly asymmetric for operation at 1.09GHz. The Y-axis has a break between 0.02 and 0.2 to clearly show the sideband intensities at both frequencies.}
\label{Fig_spectrum}
\end{figure}

Figure \ref{Fig_spectrum} shows the calculated frequency spectrum showing the intensities of the sidebands for the modulator normalized to the input pump laser intensity for operation and 177.75GHz and 1.09GHz using equation \ref{intracav_field}. The relative detuning, defined as $\frac{2\Delta}{\kappa}$, is set to 0.5. The displacement amplitude was calculated by following the derivation in \cite{partialgap}. A displacement amplification factor of $\sqrt{7}$ was accounted for displacement at 1.09GHz.

\begin{figure}[t]
\centering
\includegraphics[width=5.5in]{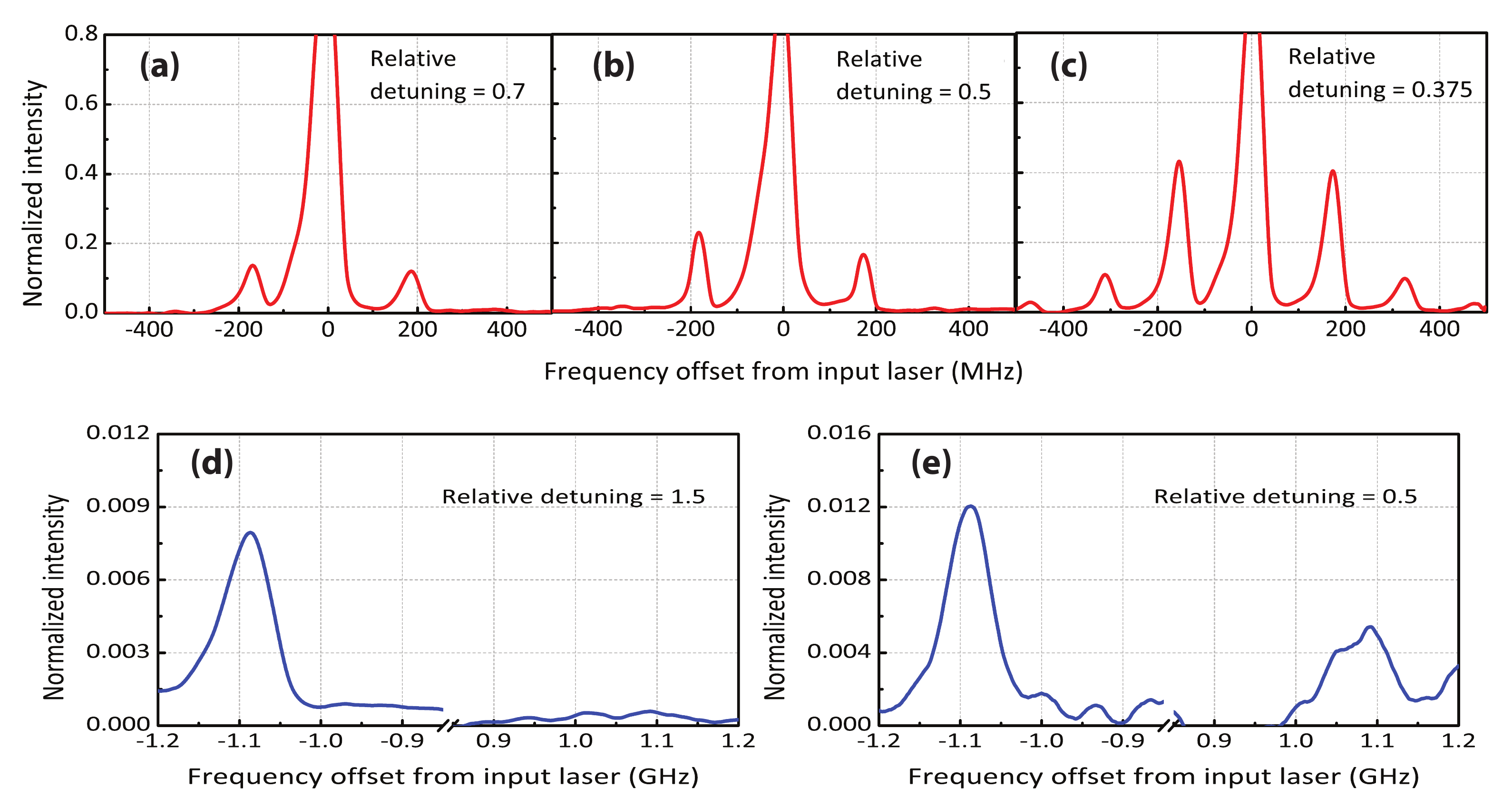}
\caption{Measured optical spectrum at output of the modulator using a scanning FP interferometer. (a)-(c) When operated at 177.75MHz for relative detuning values 0.7, 0.5 and 0.375, we observe almost symmetric lower and higher frequency sidebands in the optical spectrum on account of amplitude modulation of the laser light. A combination of 10V DC and 5dBm AC power was applied at the input electrode. The distorted shape of the laser line is an artifact of the scan. (d)-(e) The sidebands are highly asymmetric in the case of operation at 1.09GHz, which corresponds to resolved motional sidebands. Relative detuning of 1.5 corresponds to detuning on the order of the mechanical resonant frequency, resulting in maximum asymmetry. A combination of 20V DC and 8dBm AC power was applied at the input electrode. The free spectral range (FSR) of the FP interferometer used here is 1.5GHz, which leads to aliasing from adjacent sweeps and hence we only show the region of interest around ±1.09GHz.}
\label{sideband_fig}
\end{figure}

Figure \ref{sideband_fig} shows measured sideband intensities for the two mechanical modes measured at different relative detuning values. Panels (a)-(c) show recorded optical sidebands at 177.75MHz. As expected, the asymmetry in the sidebands is very small for this mechanical resonant mode of the opto-mechanical cavity, on account of unresolved motional sidebands. A combination of 10V DC and 5dBm AC power was applied at the input electrode of the modulator. Maximum asymmetry is observed for a relative detuning value of 0.5, which corresponds to the half maximum point of the optical cavity resonance. This is in sharp contrast to the highly asymmetric intensities observed in case of the mechanical resonant mode at 1.09GHz as shown in panels (d)-(e). A combination of 20V DC and 8dBm AC power was applied at the electrodes in this case. The asymmetry confirms the co-existence of AM and FM in the light field at the output of the modulator. The apparent mismatch in frequency of the positive and negative frequency sidebands is an artifact of the scan rate of the Fabry-Perot interferometer. It is possible to realize perfect FM with no AM by setting the detuning of the laser with respect to the cavity to zero, but this requires additional feedback loops, such as the Hansch-Couillaud technique \cite{ressb}. The frequency modulation factor (modulation index) for an opto-mechanical cavity was defined in equation \ref{intracav_field} as $\beta = \frac{g_{om}}{\Omega_{0}}$. The measured frequency modulation factor for our modulator operated at 1.09GHz is 0.067, which corresponds to a greater than 67X improvement over earlier reported numbers for silicon nanobeams \cite{chan}. This large modulation factor is significant for efficiently generating multiple laser lines in silicon.

\section{Conclusions}

In conclusion, we demonstrate frequency modulation at 1.09GHz of a continuous wave laser at 1,564nm using a coupled opto-mechanical silicon resonator. The monolithic integrated design augurs well for CMOS compatible processing of such modulators. This work constitutes the first demonstration of electro-acoustically generating photons of different colors from a single pump laser in silicon. The acoustic-actuation mechanism in the modulator presented here is narrowband, and governed by the mechanical quality factor. Our modulator design generates multiple laser lines, with a frequency separation governed precisely by the mechanical resonant frequency. Scaling the modulator to an internal dielectric transduced resonator design \cite{dana} will enable generation of laser lines at higher frequencies in the range of 25-50GHz for applications in dense-WDM networks.   

\section*{Acknowledgements}
The authors wish to thank Suresh Sridaran for initial fabrication of devices and help with experimental setup. We also thank Dr. Andrey Matsko of OEWaves Inc. for helpful discussions and feedback on the experiments. This work was supported by DARPA/MTO's ORCHID program and Intel Academic Research Office. The devices were fabricated at the Cornell NanoScale Science and Technology Facility. 


\bibliographystyle{IEEEtran}



\end{document}